\documentclass[14pt,a4paper]{article}
\usepackage{Packages}

\begin{document}

\vspace{0.5 cm}

{\centering
	{\bfseries\large DERIVATION OF THE THERMAL CONDUCTIVITY IN A LATENT THERMAL ENERGY STORAGE UNIT FOR USE IN SIMPLIFIED SYSTEM MODELS}
	
	\vspace{0.5 cm}
	
	Lauritz Zendel$^{1*}$, Chiara Springer$^{1}$, Frank Dammel$^{1}$, Peter Stephan$^{1}$\\ [0.2 cm]
	$^{1}$Institute for Technical Thermodynamics, Technical University of Darmstadt, Darmstadt, Germany\\ [0.2 cm]
	\textit{*Corresponding Author: zendel@ttd.tu-darmstadt.de}\\ 
	\par}
\vspace{0.5 cm}

{\bfseries \centering{ABSTRACT} \par }
Latent Thermal Energy Storages (LTES) can store thermal energy in a narrow temperature range. Therefore, they are favorable for integration into Rankine-based Carnot Batteries, heat pumps and heat engines. For the design and optimization of such systems, numerical simulations based on accurate system models are highly desirable. However, physical phenomena such as natural convection in LTES units cannot be modeled directly in transient system models. Simplified models are required. Therefore, the objective of this work is to derive simplified LTES unit models for use in system models.

While an LTES is usually modeled with a fixed temperature level in stationary simulations, in transient simulations the state of charge of the LTES influences its temperature profile. The temperature profile in turn depends on the geometry of the LTES unit. Therefore, the geometry must be taken into account in order to model the transient behavior of an LTES unit with sufficient accuracy.

The LTES unit under investigation has a shell and tube heat exchanger structure, using hexagonal finned tubes. The phase change material (PCM), which stores thermal energy by changing phase from solid to liquid, is located between the fins and in the space between the finned tubes. Aluminum fins are used. They have a high thermal conductivity and thus compensate for the low thermal conductivity of the sodium nitrate used as the PCM. The interaction between the fins and the PCM is complex and cannot be described analytically.

Therefore, a numerical approach can be used to gain insight into the behavior of the LTES unit. In order to transfer the results of a complex numerical model to a simplified model where fins and PCM are not considered individually, the effective thermal conductivity of a single finned tube can be used to approximate the performance of the LTES unit.

In this study, a finite element model of a section with a single finned tube is developed using the COMSOL Multiphysics software. 
The effective thermal conductivity of the system is determined by varying the effective thermal conductivity of both the liquid and the solid PCM in a simplified model and comparing the resulting average temperature and heat flow curves with reference cases based on a more complex modeling approach.
The results can serve as model input for simplified system models of heat pumps, heat engines or Carnot Batteries, among others.

\section{INTRODUCTION}
\begin{figure}[h]
	\centering
	\includegraphics[width=0.5\textwidth]{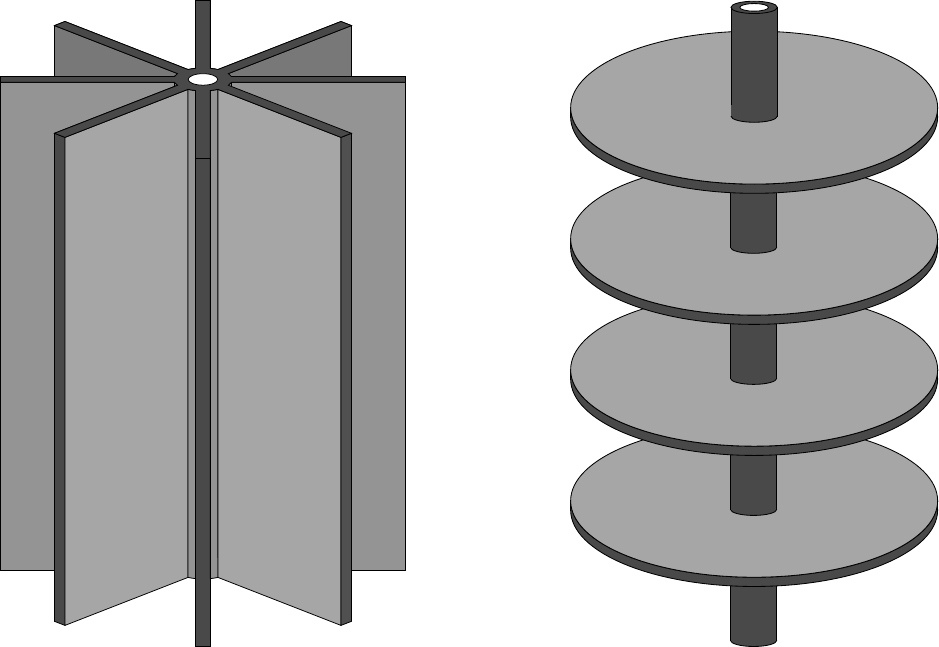}
	\caption{LTES with longitudinal fins (left) and annular fins (right)}
	\label{fig:LTES_longitudinal_and_annular}
\end{figure}

Energy storage can play an important role in energy systems. 
Thermal energy storage can be used, for example, for waste heat recovery (\cite{vanderHoeven.2014}), in buildings, in district heating networks, in power generation and in heat exchangers (\cite{Ali.2024}). In contrast to the already well-developed sensitive thermal energy storage systems, Latent Thermal Energy Storage (LTES) systems can have a higher energy density and enable efficient operation due to their narrow temperature range (\cite{Steinmann.2022}). There are different types of LTES. Active systems, encapsulation and extending the heat transfer surface are some of the possible designs (\cite{Steinmann.2022}). The latter can be further divided into longitudinal and annular finned shell-and-tube heat exchangers (\cite{Steinmann.2022}), as illustrated in Figure \ref{fig:LTES_longitudinal_and_annular}.

The influence of longitudinal fins was investigated by \cite{Kirincic.2024}. A distinction was made between charging, discharging and a combination of both processes. Two optimizations were carried out. In one, the average latent heat flow and in the other, the thermal effectiveness of the LTES unit was used as the target value. As part of the optimization, the maximum permissible aspect ratio was selected for both target values, while the thickness and number of fins took on an optimum value between the upper and lower limits. Thereby, a potential for improvement compared to an existing LTES unit could be demonstrated.

A comparison of longitudinal and annular fins was carried out by \cite{Tiari.2021b}. With the same fin volume, two longitudinal and two annular fin arrangements were experimentally compared with a configuration without fins in terms of their effects on the charging and discharging duration. It was shown that a higher number of fins, which were thinner due to the restriction of the same volume, enabled a higher thermal power and thus lower charging and discharging durations. The lowest durations were achieved with the configuration with the thin longitudinal fins.

\cite{Tiari.2021}, also investigated annular fins numerically, maintaining the requirement that the fin volume of each configuration must be identical. In addition to a constant fin length, configurations with linearly increasing and decreasing fin lengths in the axial storage direction were also investigated. It turned out that in terms of the duration for a charging and subsequent discharging cycle, the best results can be achieved with the configurations that have a constant fin length. Again, the configuration with thinner fins is better than the one with fewer thicker ones. If only the charging process is considered, the configurations in which the fins at the lower end of the LTES unit have a larger radius are superior, as they favor the positive influence of natural convection. This shows that natural convection can certainly have a relevant influence.

Nevertheless, natural convection is usually neglected in modeling in order to reduce the computational effort (\cite{Steinmann.2022}). While this effect has a negligible influence during solidification, it increases heat transfer during the melting process (\cite{Steinmann.2022}). 

The study by \cite{Vogel.2019}, focuses on the influence of natural convection on heat transfer during the charging of an LTES unit. Three branched axial configurations and one with annular fins were investigated numerically. With Ansys Fluent, 3D simulations were carried out, in order to evaluate the temporal progressions of the heat flux and the liquid phase fraction. A convective enhancement factor, which is the ratio of the heat flux taking into account natural convection to the heat flux without it, is plotted and evaluated over the liquid phase fraction. The convective enhancement factors initially increase with increasing liquid phase fraction and, depending on the configuration, show a maximum at different liquid phase fractions, which is above \qty{60}{\percent} in all cases investigated, often even close to or above \qty{90}{\percent}. For the geometry with annular fins, a maximum convective enhancement factor of \num{1.6} is achieved, although the value is significantly lower over a wide range of the liquid phase fraction. With longitudinal fins, the maximum value ranges from \num{1.1} to \num{4.7} and thus varies significantly depending on the configuration investigated.

The study by \cite{Roman.2024}, is also aimed at possible model simplifications. The LTES unit under investigation consists of several rectangular slabs filled with PCM, with air gaps between them. The influence of the thermal conductivity of the PCM and the possibility of a simplified modeling of the temperature-dependent phase change enthalpy are examined. The thermal conductivity has a small effect on the two investigated quantities, the heat flux and the outlet temperature of the heat transfer fluid air, provided that the thermal resistance due to the heat transfer from the heat transfer fluid to the pipe wall dominates the thermal resistance due to heat conduction in the PCM. The assumption of a different heat transfer fluid (HTF), a constant heat flux or a constant wall temperature could therefore significantly change the results, as the thermal resistance of HTF to the pipe is significantly reduced or completely eliminated. The interrelation between the heat transfer coefficient of the HTF to the pipe and the relevance of the thermal conductivity of the PCM is also assumed by \cite{Tehrani.2016}. As far as the temperature-dependent phase change enthalpy is concerned, it turned out that a linear relationship in the phase change interval may be sufficient (\cite{Roman.2024}). 

In the case of shell and tube heat exchangers, the approach of considering fins and PCM together is also pursued. For example \cite{Parry.2014}, scale the effective thermal conductivity of the PCM in a 1D model in order to obtain the results of a 3D model. The associated parameter is selected depending on the existing geometry and thus depending on whether and which fins are present. In the study by \cite{Waser.2018}, an analytical calculation of the effective thermal conductivity is carried out, which results from a parallel arrangement of PCM and fins. PCM and fins are weighted according to their volume fraction. The thermal conductivity obtained in this way is compared with that from a calibration, whereby similar results are achieved using the 1D model applied.

In contrast to the assumption of an exclusively parallel arrangement of PCM and fins for calculating the effective thermal conductivity, some studies introduce factors that weight the effective thermal conductivity of a parallel and a serial arrangement. \cite{Tay.2014}, derived an empirical equation using a numerical model to determine a factor that weights the thermal resistances of a parallel and serial arrangement, whereby only thermal conduction is taken into account for the calculation. The factor varies greatly depending on the fins used per meter and the thermal conductivity of the PCM employed. Based on transient simulations, the study by \cite{Dietrich.2017}, determines a factor of \num{0.45} for the weighting of the thermal conductivity of the parallel arrangement as the factor that provides the best agreement between simulation results and experimental data. For annular fins, \cite{Vogel.2020}, determined a factor of \num{0.8} for the weighting of the parallel arrangement, whereas a factor of \num{0.7} was determined for longitudinal fins. 

It becomes clear that the factors determined, although they are not defined identically and therefore deviations between the values are to be expected, differ considerably from one another. 
In this study, the effective thermal conductivity of an annular finned tube with PCM between the fins is therefore examined using a detailed COMSOL model. The effective thermal conductivity of PCM with fins is determined for both the liquid and the solid phase of the PCM, which best approximates a detailed model that serves as a reference case.  
The results obtained in this way can be used for system simulations and complement previous studies aimed at estimating the effective thermal conductivity.

\section{MODELING}

The modeling is carried out with the software COMSOL Mulitphysics Version 6.2. A 2D axisymmetric computational domain is assumed.
Structured rectangular grids with linear shape functions for the temperature, the velocity components and the pressure field are applied.

\subsection{Modeled section and material properties}

\begin{figure}[ht]
	\centering
	\includegraphics[width=\textwidth]{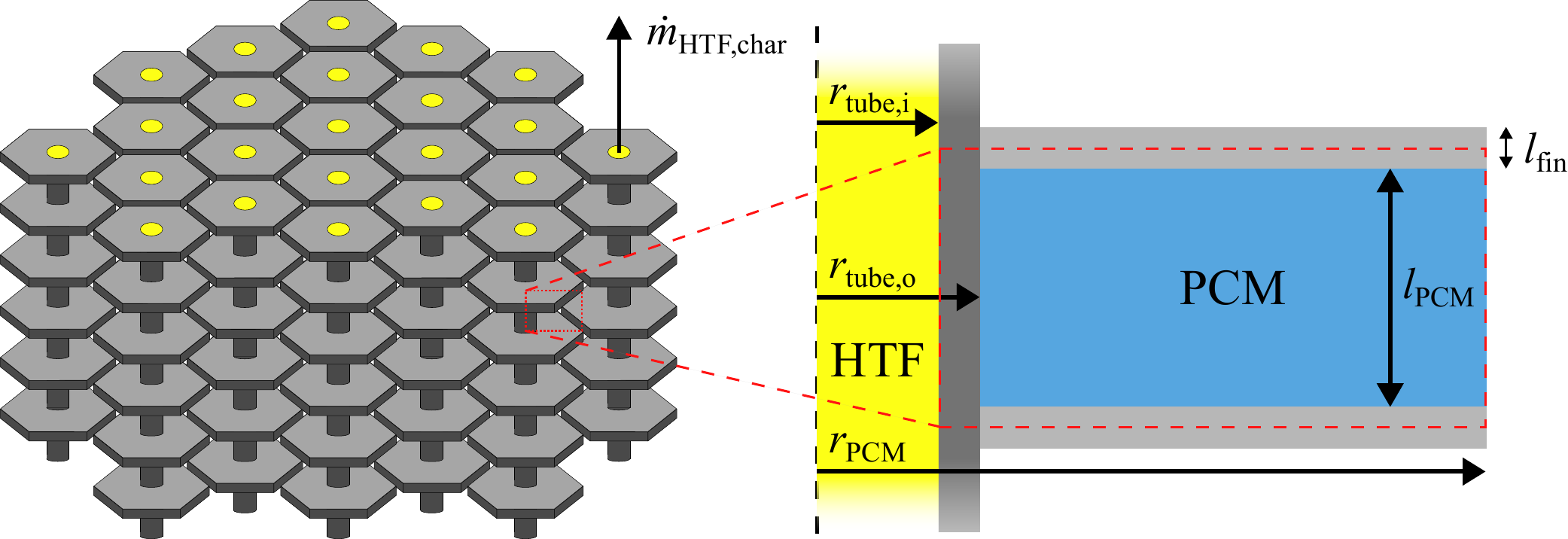}
	\caption{Section of the LTES unit and detailed view, with the modeled representative section highlighted by a red dashed line. The modeled representative section is assumed to be annular
	}
	\label{fig:LTES_geometry_descriptions}
\end{figure}

The LTES unit under investigation has the geometry of a vertical shell-and-tube heat exchanger. Each individual tube has hexagonal fins. In this study, a representative section of such a tube is modeled in order to approximate the overall behavior of the LTES unit. For this purpose, the hexagonal fins are modeled as annular fins. Figure \ref{fig:LTES_geometry_descriptions} shows a section of the LTES unit examined and the modeled representative section.

The tube in which the HTF flows is made of steel, the fins of aluminum and sodium nitrate ($\mathrm{NaNO}_{3}$) is chosen as PCM. 
The dimensions of the tube are given in Table \ref{tab:Tube_dimensions} and are derived from the dimensions of the laboratory test rig by \cite{Laing.2009}, given in \cite{Dietrich.2017}. The experimental data from \cite{Laing.2010}, are used for validation. Similar dimensions were also used for a \qty{680}{\kWh} prototype that was installed in Spain (\cite{Laing.2010}). 
The material properties are listed in Table 2, which also contains the data of the HTF Therminol VP-1 used in the validation and backtesting case.
If necessary, the material properties were interpolated to a temperature of \qty{306}{\degreeCelsius}.

\begin{table}[H]
	\centering
	\caption{Finned tube dimensions in mm}
	\begin{tabular}{cllll}
		\toprule         $r_\mathrm{tube,i}$ & $r_\mathrm{tube,o}$ & $r_\mathrm{PCM}$ & $l_\mathrm{PCM}$ & $l_\mathrm{fin}$ \\
		\midrule 
		5 & 6 & 50 & 10  & 1 \\
		\bottomrule
	\end{tabular}
	\label{tab:Tube_dimensions}
\end{table}

To avoid mass changes of $\mathrm{NaNO}_{3}$ in the calculation area, the average value of $\rho_\mathrm{NaNO_3,av}~=~\qty{2014}{\kg\per\cubic\meter}$ from the density of the liquid phase of $\rho_\mathrm{NaNO_3,l}~=~\qty{1908}{\kg\per\cubic\meter}$ (\cite{Bauer.2012}) and the solid phase of $\rho_\mathrm{NaNO_3,s}~=~\qty{2120}{\kg\per\cubic\meter}$ (\cite{Dietrich.2017}) was used in both phases.
Temperature-independent values were assumed for the thermal conductivity and dynamic viscosity in the solid phase. The following temperature-dependent thermal conductivity from \cite{Nagasaka.1991}, was assumed for the liquid PCM: 

\begin{equation}
	\lambda_\mathrm{NaNO_3,l} = \left(0.6203 - 1.8 \times 10^{-4} \cdot \frac{T}{\mathrm{K}} \right) \qty{}{\watt\per\meter\per\kelvin}
	\label{eq:lambda_NaNO3_liq}
\end{equation}

The temperature-dependent dynamic viscosity is taken from \cite{Nunes.2006}, where $\eta_\mathrm{base} = \qty{3.033}{\milli\pascal\second}$ is to be used:

\begin{equation}
	\eta_\mathrm{NaNO_3,l} = \eta_\mathrm{base} \cdot \mathrm{e}^{\left(26.689 - \frac{97.54}{\frac{T}{\qty{580}{\kelvin}}} + \frac{112.5}{\left(\frac{T}{\qty{580}{\kelvin}}\right)^2} - \frac{41.7}{\left(\frac{T}{\qty{580}{\kelvin}}\right)^3}
		\right)} \qty{}{\kg\per\meter\per\second}
	\label{eq:eta_NaNO3_liq}
\end{equation}

\begin{table}[h!]
	\centering
	\caption{Material properties with data taken or linearly interpolated [i] to the temperature \qty{306}{\degreeCelsius} from \cite{Ullrich.2019}, \cite{Dietrich.2017}, \cite{Bauer.2012},  \cite{Jriri.1995},    \cite{Janz.1967}, and \cite{EastmanChemicalCompany.2022}}
	\begin{tabular}{ccccc}
		\toprule  
		Property & Steel & Aluminum & $\mathrm{NaNO}_{3}$ & Therminol VP-1 \\ 
		\midrule
		$\rho \text{ in } \qty{}{\kg\per\meter\cubed}$ & \num{7980}  & \num{2710}  & \num{2014} & \num{810.4} [i] \\ 
		\midrule 
		$c \text{ in } \qty{}{\joule\per\kg\per\kelvin}$ & \num{531.8} [i] & \num{1034.88} [i] & \num{1658} & \num{2330.2} [i]\\ 
		\midrule
		$\lambda \text{ in } \qty{}{\watt\per\meter\per\kelvin}$ & \num{18.12} [i] & \num{231.64} [i] \ & Eq. \ref{eq:lambda_NaNO3_liq} / \num{0.7}  & \num{0.0953} [i]\\
		\midrule
		$\eta \text{ in } \qty{}{\kg\per\meter\per\second}$  & - & - & Eq. \ref{eq:eta_NaNO3_liq} / $\eta_\mathrm{base} \times 10^{6}$& $2.15 \times 10^{-4}$ [i]\\ 
		\midrule
		$t_\mathrm{PC} \text{ in } \qty{}{\degreeCelsius}$ & - & - & \num{306} & -  \\ 
		\midrule
		$\Delta T_\mathrm{PC} \text{ in } \qty{}{\kelvin}$ & - & - & \num{5} & - \\ 
		\midrule
		$h_\mathrm{PC} \text{ in } \qty{}{\kJ\per\kg}$ & - & - & \num{178} & - \\ 
		\midrule 
		$\beta \text{ in } \qty{}{\per\K}$ & - & - & $7.15 \times 10^{-4}$ & - \\ 
		\bottomrule
	\end{tabular}
	\label{tab:Material_properties}
\end{table}

Four different models are used for the study.
These are shown in Figure \ref{fig:LTES_model_section} and are described in the following.

\subsection{Validation model}
\begin{figure}[h]
	\centering
	\begin{subfigure}[b]{0.40\textwidth}
		\centering
		\begin{overpic}[width=\textwidth]{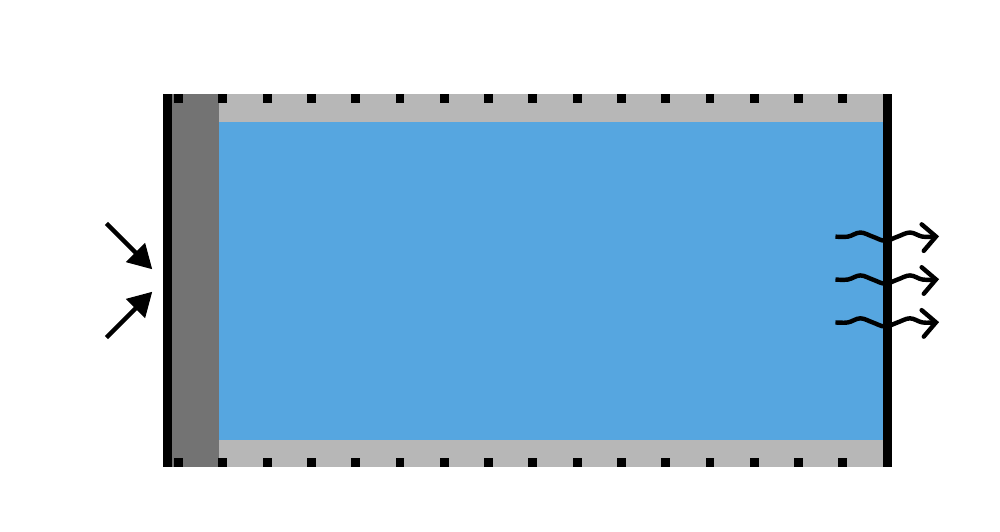}
			\put (6.5,17) {$\alpha$} 
			\put (-12,32) {$t_\mathrm{HTF,av,exp}$}
			\put (25,1) {\small Periodic boundary condition}
			\put (25,46) {\small Periodic boundary condition}
			\put (95,23.5) {$\dot{Q}_\mathrm{loss}$}
		\end{overpic}
		\caption{Modeled representative section for \\ validation}
		\label{subfig:LTES_model_section_validation}
	\end{subfigure}
	\hspace{0.09\textwidth}
	\begin{subfigure}[b]{0.40\textwidth}
		\centering
		\begin{overpic}[width=\textwidth]{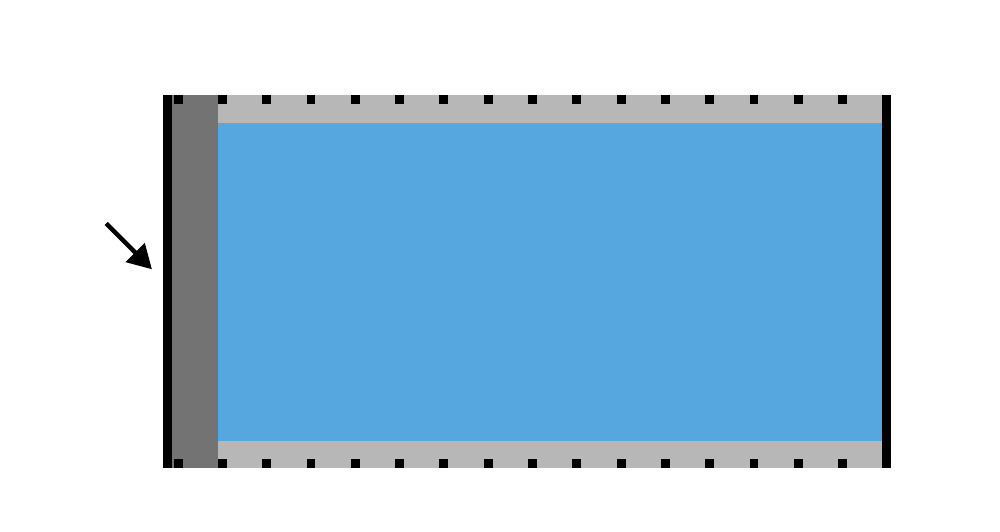}
			\put (1,32) {$t_\mathrm{tube}$}
			\put (25,1) {\small Periodic boundary condition}
			\put (25,46) {\small Periodic boundary condition}
			\put(91,15){\rotatebox{90}{\small adiabatic}}
		\end{overpic}
		\caption{Modeled representative section for effective thermal conductivity investigation - reference case}
		\label{subfig:LTES_model_reference_case}
	\end{subfigure}
	
	\vspace{0.2cm}
	
	\begin{subfigure}[b]{0.40\textwidth}
		\centering
		\begin{overpic}[width=\textwidth]{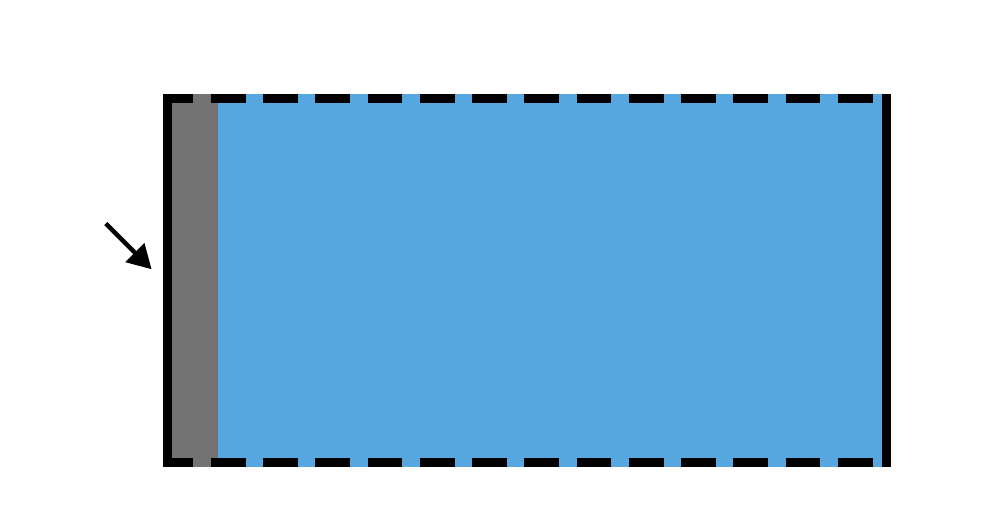}
			\put (1,32) {$t_\mathrm{tube}$}
			\put (22,1) {\small Symmetry boundary condition}
			\put (22,46) {\small Symmetry boundary condition}
			\put(91,15){\rotatebox{90}{\small adiabatic}}
		\end{overpic}
		\caption{Modeled representative section for effective thermal conductivity investigation - conductivity only}
		\label{subfig:LTES_model_lambda_determination}
	\end{subfigure}
	\hspace{0.09\textwidth}
	\begin{subfigure}[b]{0.40\textwidth}
		\centering
		\begin{overpic}[width=\textwidth]{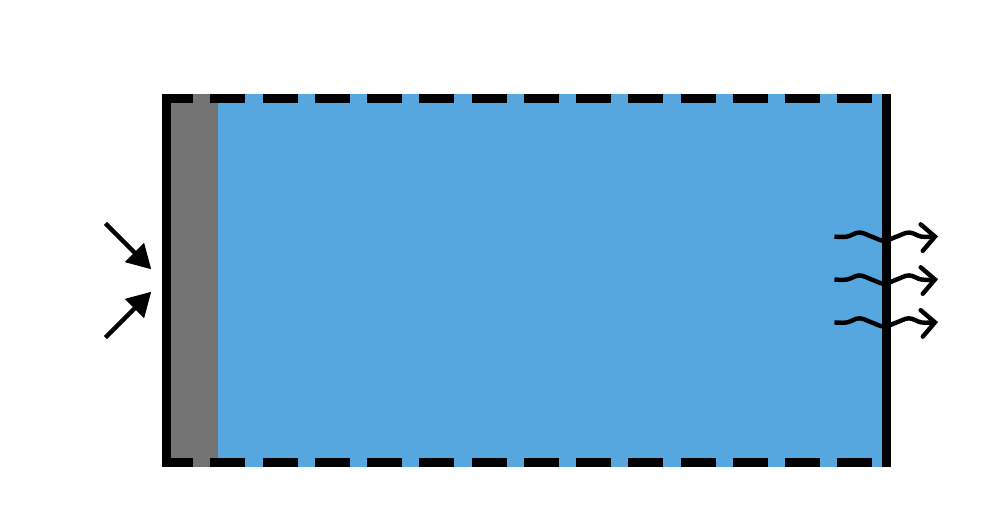}
			\put (6.5,17) {$\alpha$} 
			\put (-12,32) {$t_\mathrm{HTF,av,exp}$}
			\put (22,1) {\small Symmetry boundary condition}
			\put (22,46) {\small Symmetry boundary condition}
			\put (95,23.5) {$\dot{Q}_\mathrm{loss}$}
		\end{overpic} 
		\caption{Modeled representative section for backtesting - conductivity only  \\  \, }
		\label{subfig:LTES_model_backtesting}
	\end{subfigure}
	\caption{Modeled representative section for validation (a), for effective thermal conductivity investigation - reference case (b), for effective thermal conductivity investigation - conductivity only (c) and for backtesting - conductivity only (d).
	}
	\label{fig:LTES_model_section}
\end{figure}
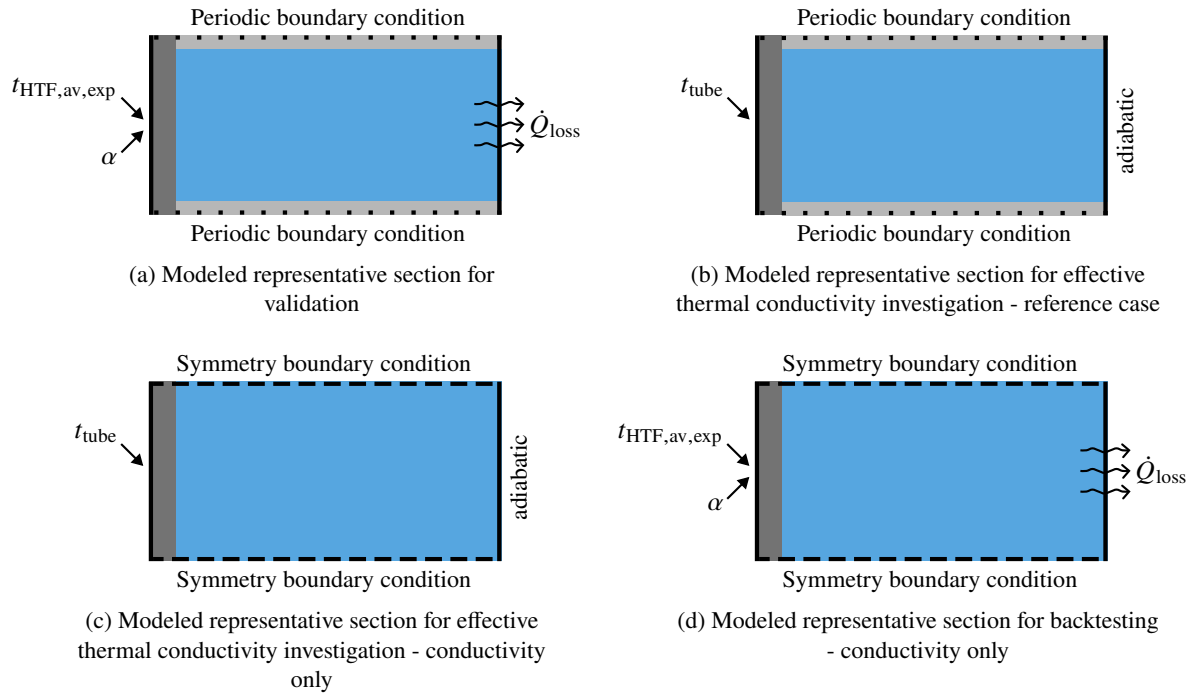

The model for the validation case is shown in Figure \ref{subfig:LTES_model_section_validation}. Convective heat transfer is present on the inside of the tube carrying the HTF.
The input and the output temperature of the HTF as well as the average PCM temperature were visually reconstructed using the online tool WebPlotDigitizer (\cite{Rohatgi.2023}). While the experimental average temperature of the HTF is specified in the model the experimental average PCM temperature serves as validation value.
A flow velocity $v_\mathrm{HTF} = \qty{1.97}{\meter\per\second}$ was selected according to the data from \cite{Laing.2009}.
The correlations of \cite{Gnielinski.2019}, were used to calculate the Nusselt number, assuming a fully developed turbulent pipe flow. A heat transfer coefficient $\alpha = \qty{4060.3}{\watt\per\meter\squared\per\kelvin}$ was calculated in this way.

A heat loss flow is given on the outside of the pipe containing the PCM. The value $\dot{Q}_\mathrm{loss} = \qty{0.4931}{\watt}$, which is based on the losses reported in \cite{Laing.2009}, and the ratio of the section considered for modeling compared to the total length of all storage pipes, is specified.
Detailed modeling is not required for transfer to a real storage tank, as the heat losses can be neglected if the real tank is sufficiently large. 

Periodic boundary conditions are assumed for the top and bottom of the calculation area.

In COMSOL Multiphysics, the Heat Transfer in Solids and Fluids Interface and the Laminar Flow Interface are coupled with each other using Nonisothermal Flow. 
In this way, both the heat conduction in the entire calculation area and the flow of the PCM, which is assumed to be an incompressible Newtonian fluid for which conservation of mass, momentum and energy apply (\cite{COMSOLAB.2024}), can be modeled. This enables the modeling of natural convection.
The underlying equations can be found in the documentation (\cite{COMSOLAB.2024}).

\begin{figure}[ht]
	\centering
	\includegraphics[width=\textwidth]{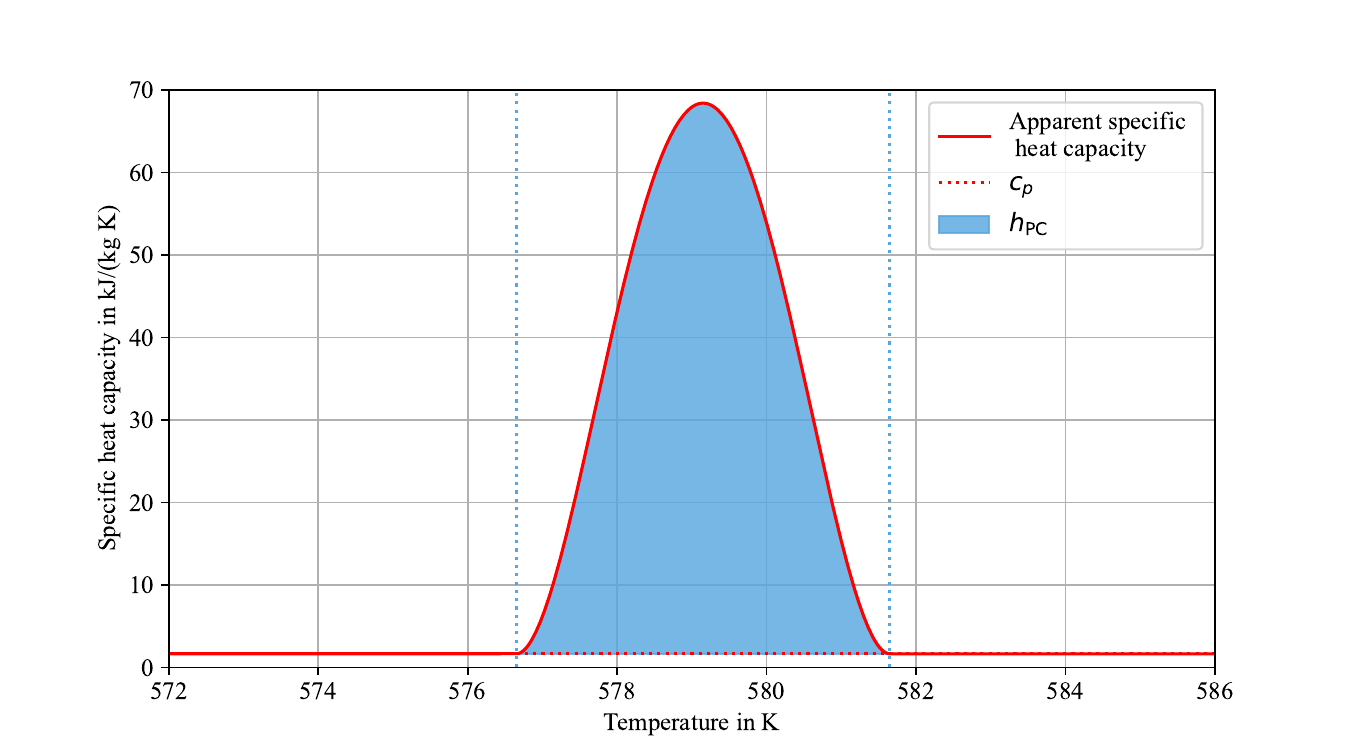}
	\caption{Distribution of the phase change enthalpy $h_\mathrm{PC}$ in the model with the apparent heat capacity method. The blue dotted lines delimit the phase change interval}
	\label{fig:fig_apparent_heat_capacity}
\end{figure}

A phase change material is added as a liquid for modeling the PCM. This contains the method of apparent heat capacity. The phase change temperature, the phase change interval and the phase change enthalpy must be specified, which can be found in Table \ref{tab:Material_properties}. 
The resulting apparent specific heat capacity as a function of temperature, the distribution of the phase change enthalpy and the temperature interval of this distribution can be seen in Figure \ref{fig:fig_apparent_heat_capacity}. 
The coefficient of thermal expansion is utilized to take natural convection into account. Since the density is specified as constant, but a volume force dependent on the temperature is defined, the Boussinesq approximation is applied in this way. 

\subsection{Reference case model}

The reference case model is shown in Figure \ref{subfig:LTES_model_reference_case}.
The first step for further investigations is to define reference cases so that studies can be carried out independently of the validation case. The model from the validation is modified in such a way that a temperature is specified instead of the temperature of the HTF and a heat transfer coefficient. 
In addition, the losses are set to zero, in accordance with the fact that heat losses become negligible with large storage capacities.

\subsection{Models with PCM only}
In order to be able to integrate a resulting effective thermal conductivity of the investigation into a system model, it is necessary to reduce the complexity of the model. Therefore, in the following, the fins and the PCM are not modeled separately, but exclusively by the PCM, whose material properties are adjusted accordingly. 

For the investigation of the effective thermal conductivity (Figure \ref{subfig:LTES_model_lambda_determination}), a constant temperature is set on the inside of the tube carrying the HTF, which would be present if the HTF were to condense or vaporize. 
An adiabatic boundary condition is defined on the outside.
Symmetry boundary conditions are assumed both at the bottom and at the top.

The volume force and thus the natural convection are not modeled. Instead, only thermal conduction of the PCM is modeled. The aluminum fins are replaced by the PCM.
To ensure comparability with the cases in which the fins are modeled, the density of the PCM is multiplied by the factor $\frac{l_\mathrm{fin}}{l_\mathrm{PCM} + l_\mathrm{fin}}$ so that the heat capacity of the PCM remains unchanged, whereas that of the aluminum fins is omitted.
This is justified by the fact that the aluminum fins are responsible for less than \qty{2.5}{\percent} of the heat capacity of the fins and PCM in the case of the largest temperature interval passed through and thus at the greatest influence of the sensible amount of heat that can be stored in the fins.

To carry out a subsequent backtesting, the model described before is applied, whereby the values from the validation are used instead of the temperature specification and the losses are taken into account again, as can be seen in Figure \ref{subfig:LTES_model_backtesting}.

\subsection{Evaluation of effective thermal conductivity}

The aim of this study is to determine the most representative effective thermal conductivity for both the liquid and the solid phase. To determine these values, reference cases are first defined. 
In these cases, a temperature of \qty{5}{\kelvin}, \qty{10}{\kelvin} and \qty{15}{\kelvin} above the melting temperature is applied to the inner wall of the tube for the charging process. 
For the subsequent discharging, temperatures are applied that are below the melting temperature by the same amounts.
These temperatures are also specified as the initial temperatures.

The values presented are typical (\qty{5}{\kelvin}, \qty{10}{\kelvin}) (\cite{Steinmann.2022}) or greater (\qty{15}{\kelvin}) than those for technical applications and therefore serve as the basis for the present study.
The phase change temperature was used as the basis. The differences between the initial temperatures and the temperatures applied on the inner wall of the tube during charging are therefore twice the temperature differences mentioned.

In the model depicted in Figure \ref{subfig:LTES_model_lambda_determination}, all values between \qty{10}{\watt\per\meter\per\kelvin} and \qty{20}{\watt\per\meter\per\kelvin} are run through in steps of \qty{1}{\watt\per\meter\per\kelvin} for both the liquid and the solid PCM applying the same temperatures as in the reference cases.  

The root mean square error (RMSE) is used to determine the combinations that show the smallest deviations from the respective reference case for the temperatures and heat flows for the charging and discharging process. 
For the calculation of the RMSE, the values of the first \qty{15}{\min} of charging and discharging are not taken into account in the case of the heat flows due to the strong spikes caused by high temperature gradients. In addition, the last \qty{5}{\min} of the charging process are excluded.

\section{VALIDATION}
\begin{figure}[ht]
	\centering
	\includegraphics[width=\textwidth]{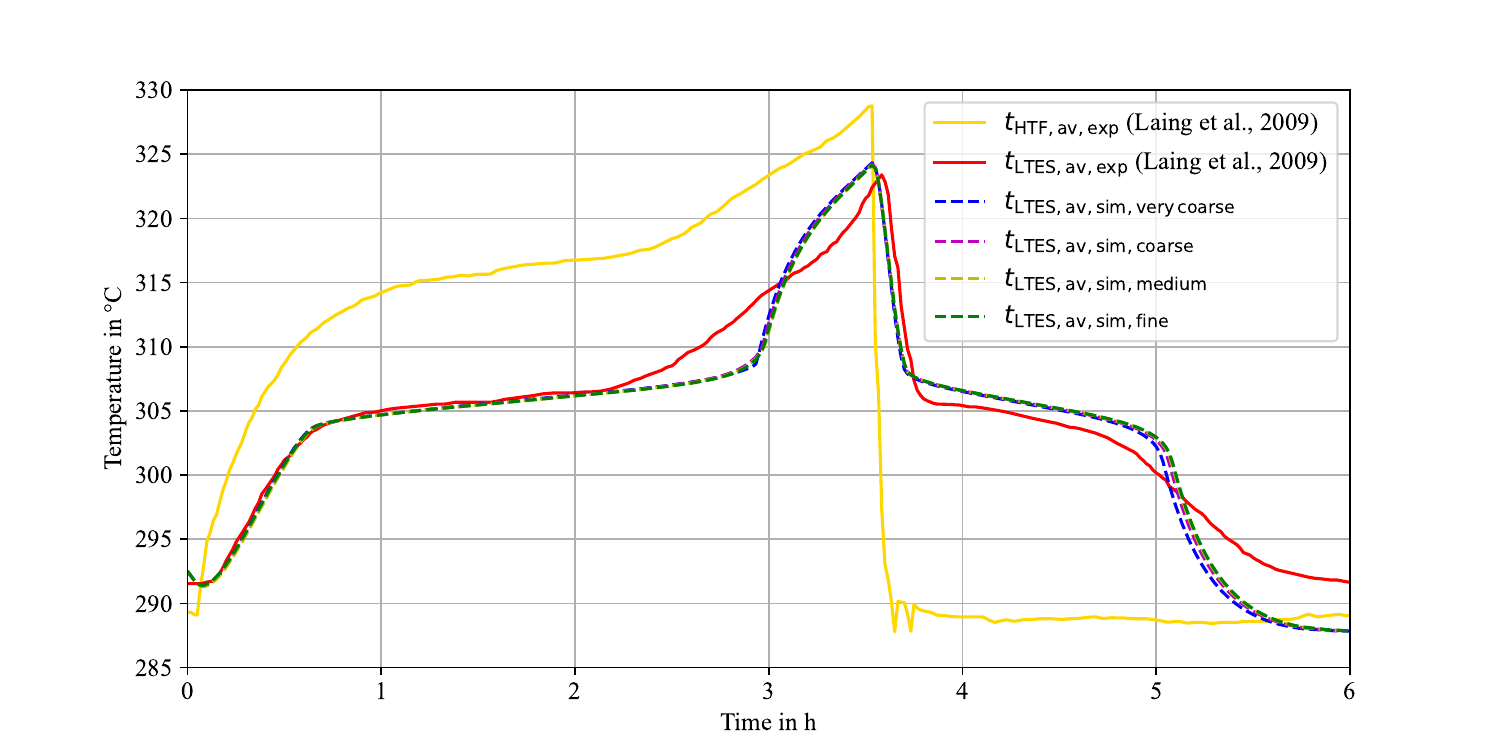}
	\caption{Experimental and simulation results of the average LTES unit temperature}
	\label{fig:Validation_temperature}
\end{figure}

The validation is based on experimental data taken from \cite{Laing.2010}. The reconstructed experimental data and the simulation results of the PCM temperature, which is calculated volumetrically averaged, is shown in Figure \ref{fig:Validation_temperature}. 
As can be seen from the figure, a grid study was carried out. 
Four structured rectangular grids with a number of \num{855} (very coarse), \num{3277} (coarse), \num{12600} (medium) and \num{49500} grid elements (fine) were examined.
The coarse grid proved to be sufficiently precise and was therefore used for further investigations.

The transition between the phase change process, which is accompanied by a flat temperature increase, and the further temperature increase of the PCM due to the absorption of sensible heat is more evident in the simulation than in the experiment. One of the reasons for this may be that only a section of the storage tank is considered in the model, whereas in the experiment the PCM may have reached different stages of melting along the direction of flow of the HTF. 
In the experiment, the average temperature can therefore rise as a result of the temperature increase of the PCM in parts of the LTES unit, while other parts of the LTES unit are still undergoing a phase change.

Furthermore, there are measurement uncertainties in the experiment that can affect both the average PCM temperature and the HTF temperatures. The latter were used for the validation case and backtesting, which means that deviations here can also affect the model.

Based on the above explanations, the model is therefore considered validated.

\section{RESULTS AND DISCUSSION}

In this chapter, the results of the effective thermal conductivity investigation and the backtesting are presented. Afterward, a comparison is made with the literature.

\begin{figure}[ht]
	\centering
	\includegraphics[width=\textwidth]{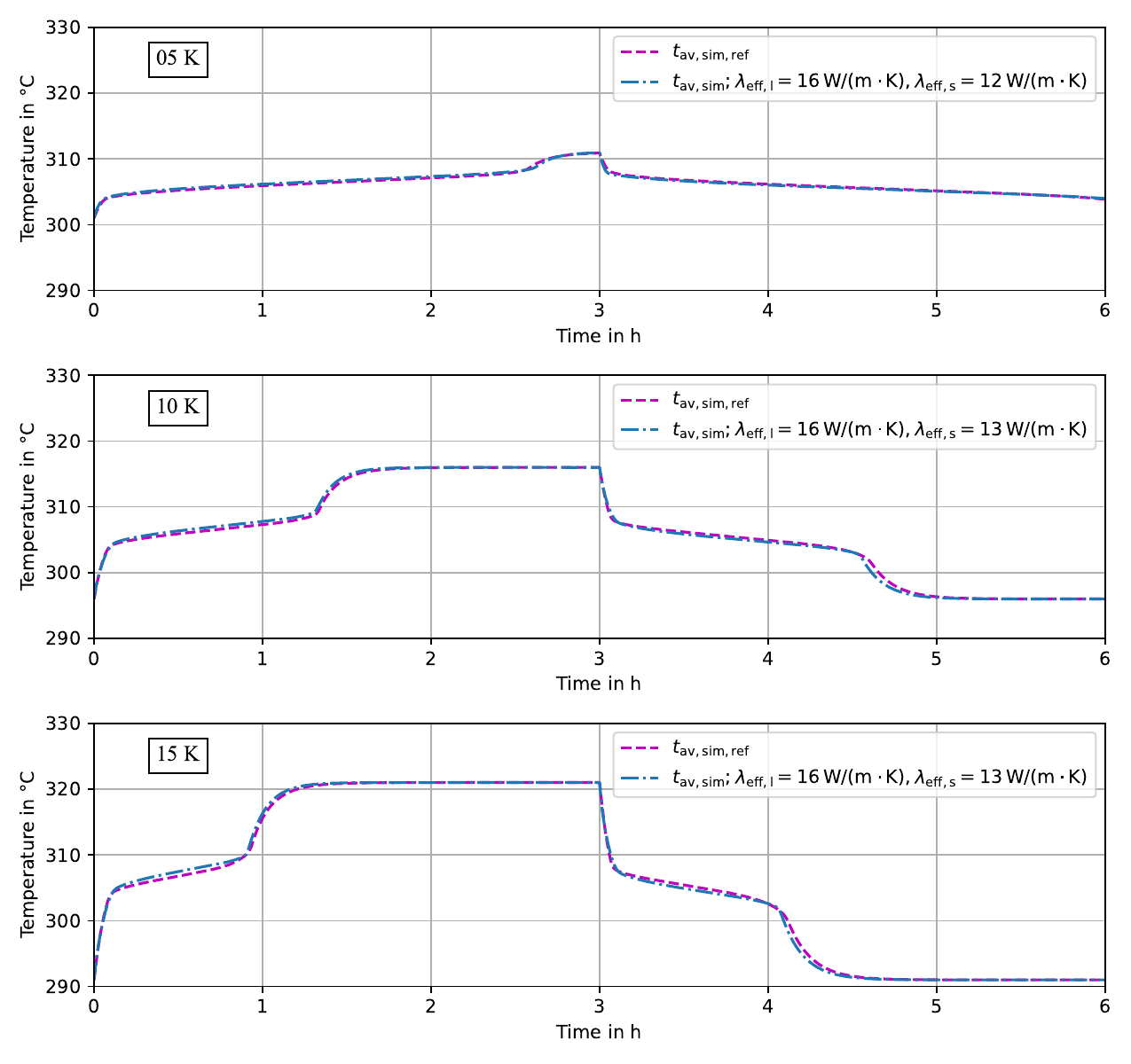}
	\caption{Average temperatures of the reference simulations and simulations with the lowest RMSE. From top to bottom: temperature difference between inner tube wall and melting temperature: \qty{5}{K}, \qty{10}{K}, \qty{15}{K}.}
	\label{fig:T_av_ref_and_opt_lambdas}
\end{figure}

\subsection{Effective thermal conductivity investigation}

\begin{figure}[h!]
	\centering
	\includegraphics[width=\textwidth]{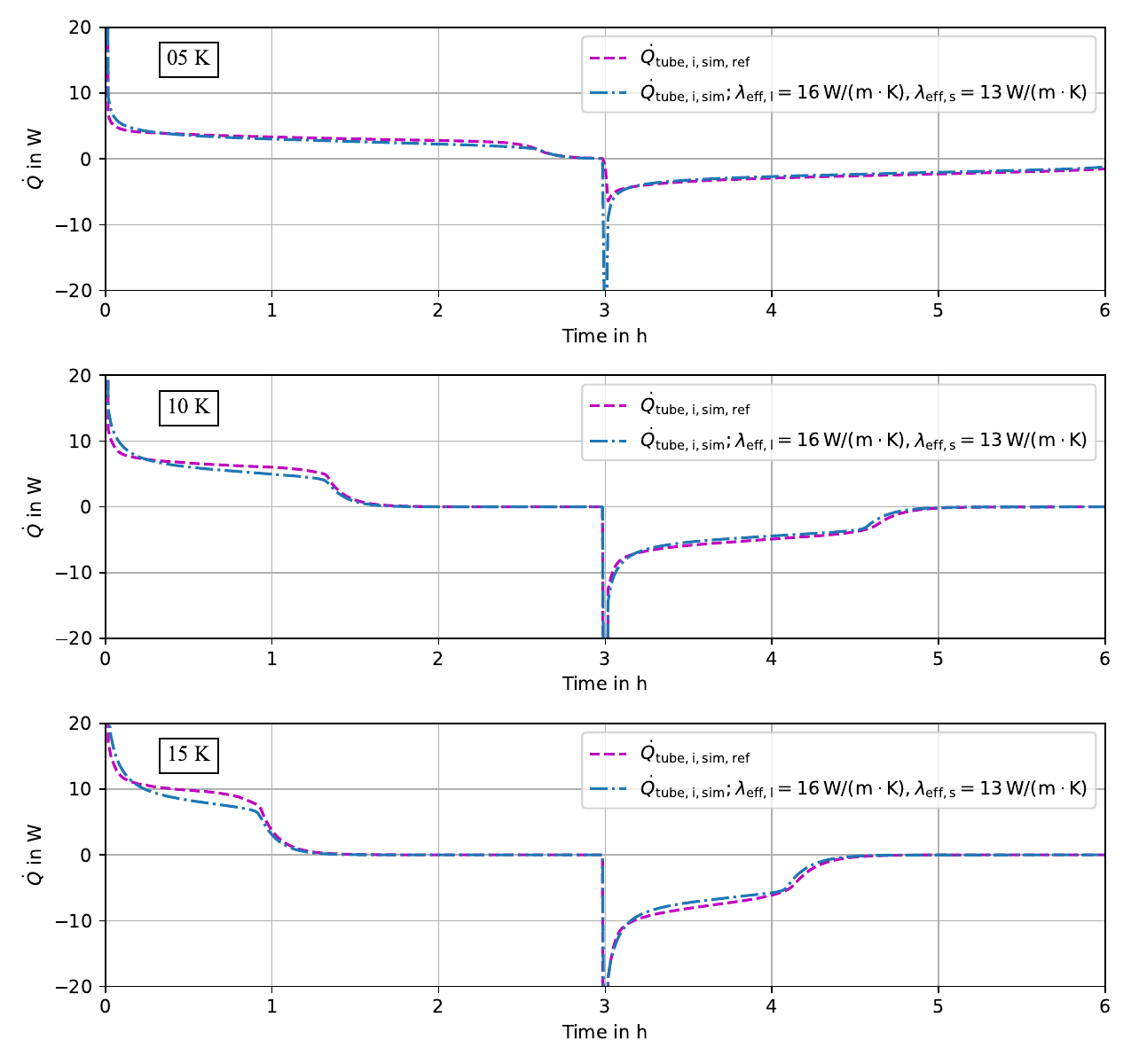}
	\caption{Heat flows on the inner wall of the tube to the modeled representative section of the reference simulations and simulations with the lowest RMSE. From top to bottom: temperature difference between inner tube wall and melting temperature: \qty{5}{K}, \qty{10}{K}, \qty{15}{K}}
	\label{fig:Q_dot_in_ref_and_opt_lambdas}
\end{figure}

For the investigation of the effective thermal conductivity, \qty{3}{\hour} of charging and \qty{3}{\hour} of discharging are specified.
The results of the reference cases and the corresponding combinations of the effective thermal conductivities of the liquid and solid PCM with the lowest RMSE with respect to the average temperature are shown in Figure \ref{fig:T_av_ref_and_opt_lambdas}.
The other \num{120} combinations resulting from the variation of the effective thermal conductivities are not shown for reasons of clarity.
For a temperature difference of \qty{5}{\kelvin}, it can be seen that the temperature specified on the inner wall of the tube has not yet been reached, both in the charging process and especially in the discharging process.
Figure \ref{fig:Q_dot_in_ref_and_opt_lambdas} shows the results of the reference cases and the combinations that have the lowest RMSE in terms of heat flow on the inner wall of the tube. 
For clarity, the additional \num{120} combinations are again not shown and the visible range of the heat flows is reduced to \qty{-20}{\watt} to \qty{20}{\watt}. Thus, the spikes in heat flow at the beginning of the charging and discharging process are not shown completely in the figure.
In accordance with the temperature curve at the temperature difference of \qty{5}{\kelvin} in Figure \ref{fig:T_av_ref_and_opt_lambdas}, it can also be recognized from the heat flow that the discharging process in particular was not yet complete, as the heat flow has not yet reached zero.

The agreement between the simulation results with the simplified model and those with the corresponding reference cases is higher for the average temperatures than for the heat flows on the inner wall of the tube. The latter show higher deviations, especially at higher temperature differences (bottom of Figure \ref{fig:Q_dot_in_ref_and_opt_lambdas}).

It can be seen that the effective thermal conductivities with the highest agreement for both cases and all three investigated temperature differences show almost exclusively identical values.
Only for the average temperature at a temperature difference of \qty{5}{\kelvin} (top of Figure \ref{fig:T_av_ref_and_opt_lambdas}) the results based on the effective thermal conductivities $\lambda_\mathrm{eff,l} = \qty{16}{\watt\per\meter\per\kelvin}$ and $\lambda_\mathrm{eff,s} = \qty{12}{\watt\per\meter\per\kelvin}$ show the highest agreement. 
In all other cases, the results using effective thermal conductivities $\lambda_\mathrm{eff,l} = \qty{16}{\watt\per\meter\per\kelvin}$ and $\lambda_\mathrm{eff,s} = \qty{13}{\watt\per\meter\per\kelvin}$ are associated with the smallest deviations from the reference cases. 
It is noticeable that the case in which the charging and discharging process was not fully completed in the simulation time is the deviating case. 

\subsection{Backtesting}

In order to investigate the applicability of the effective thermal conductivities obtained to the validation case, backtesting is carried out. The model from Figure \ref{subfig:LTES_model_backtesting} is used for this, assuming an effective thermal conductivity of \qty{16}{\watt\per\meter\per\kelvin} for the liquid PCM and \qty{13}{\watt\per\meter\per\kelvin} for the solid PCM based on the results. 

\begin{figure}[h!]
	\centering
	\includegraphics[width=\textwidth]{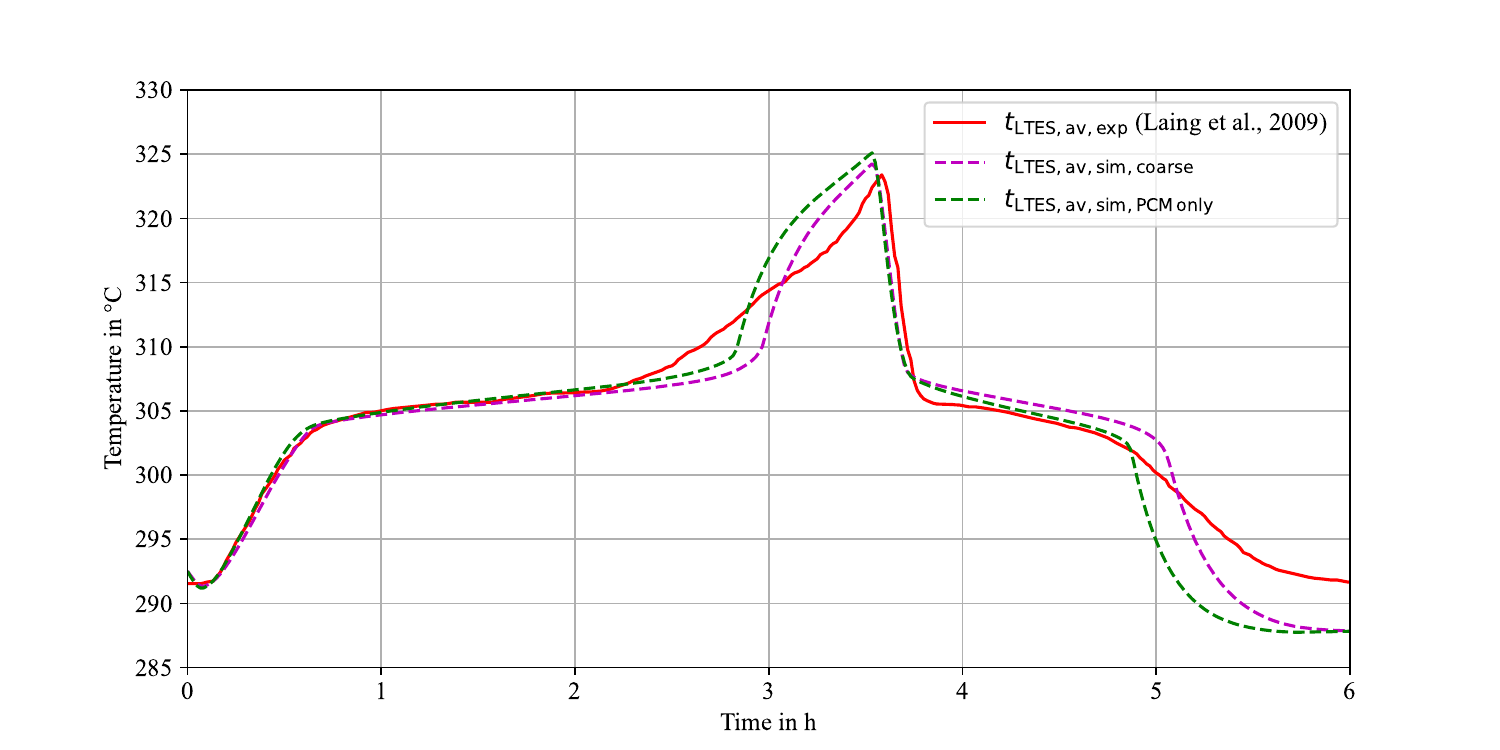}
	\caption{Experimental, validation simulation and backtesting simulation results of the average LTES unit temperature}
	\label{fig:Backtesting_temperature}
\end{figure}

Figure \ref{fig:Backtesting_temperature} illustrates the outcomes. In addition to the measured values and the described case, the validation case with the coarse grid is shown to ensure direct comparability.
In the charging process, the average temperature starts to rise earlier in the simulation that only includes the PCM. In addition, a higher maximum average temperature is reached. The average temperature also drops earlier during the discharging process and ends at approximately the same temperature as it ends with the validated model.

It is clear that the reduction of the model complexity leads to an increase in the deviation between the simulated and the measured average temperature of the LTES. Nevertheless, it can be stated that the model provides an adequate representation of the measured values in order to use the modeling approach in a system model.
This can be concluded especially against the background that even the more complex model shows recognizable deviations from the experimental data.

\subsection{Comparison to literature}

Assuming a completely parallel arrangement of fins and PCM, the resulting effective thermal conductivities for the given configuration and the existing material values would be \qty{21.56}{\watt\per\meter\per\kelvin} and \qty{21.69}{\watt\per\meter\per\kelvin} for the liquid and solid phases of the PCM, respectively. 
A value of \qty{0.55}{\watt\per\meter\per\kelvin} (\cite{Dietrich.2017}) is assumed for the thermal conductivity of the liquid PCM. 
According to the study by \cite{Vogel.2020}, which assumes a factor of \num{0.8} for the weighting of the parallel arrangement, this would result in values of \qty{17.37}{\watt\per\meter\per\kelvin} and \qty{17.51}{\watt\per\meter\per\kelvin} for the effective thermal conductivities mentioned. 
Compared to these two studies, the present investigation suggests lower effective thermal conductivities. The comparison with the weighting factor of \cite{Dietrich.2017}, which would result in effective thermal conductivities of \qty{10.03}{\watt\per\meter\per\kelvin} and \qty{10.19}{\watt\per\meter\per\kelvin}, shows that the effective thermal conductivities in the present study are higher. The effective thermal conductivities determined in this study correspond to weighting factors of approximately \num{0.735} and \num{0.585} for the liquid and solid phases respectively.

\section{CONCLUSION}
\label{sec:conclusion}

This paper presents a simplified model of an LTES unit with hexagonal fins, modeled as annular fins. 
By replacing the structure of fins and PCM with the PCM only, it is possible to significantly reduce the model complexity. By comparing simulations that consider effective thermal conductivity for both the liquid and the solid PCM individually with simulations of a validated model that accounts for the exact geometry and natural convection, approximated values for the effective thermal conductivities could be determined. These are \qty{16}{\watt\per\meter\per\kelvin} and \qty{13}{\watt\per\meter\per\kelvin} for the liquid and solid PCM. 
The comparison of the simulation results with regard to the average LTES unit temperature shows that the simplified model has higher deviations. Nevertheless, the model can be considered sufficiently precise for use in system modeling. 

In future studies, the procedure presented in this paper could be applied to configurations that are modified in terms of geometry or material. The influence of the varied parameters could be quantified in this way.

\section*{NOMENCLATURE}
\label{sec:symbole_abk}

\subsection*{Abbreviations}
\begin{tabular}{@{}p{2cm}l}
	HTF & Heat transfer fluid\\
	LTES & Latent thermal energy storage\\
	PCM & Phase change material \\
	RMSE & Root mean square error \\
\end{tabular}

\subsection*{Symbols}
\begin{tabular}{@{}p{2cm}l}
	$c$ & specific heat capacity, \qty{}{\J\per\kg\per\kelvin} \\
	$h$ & specific enthalpy, \qty{}{\kJ\per\kg} \\
	$l$ & length, \qty{}{\mm} \\
	$\dot{m}$ & mass flow rate, \qty{}{\kg\per\second} \\
	$\dot{Q}$ & heat flow, \qty{}{\watt} \\
	$r$ & radius, \qty{}{\mm} \\
	$t$ & temperature, \qty{}{\degreeCelsius} \\
	$T$ & temperature, \qty{}{\kelvin} \\
	$\Delta T$ & temperature difference, \qty{}{\kelvin} \\
	$v$ & velocity, \qty{}{\meter\per\second} \\
	$\alpha$ & heat transfer coefficient, \qty{}{\watt\per\meter\squared\per\kelvin} \\
	$\beta$ & coefficient of thermal expansion in \qty{}{\per\kelvin} \\
	$\eta$ & dynamic viscosity, \qty{}{\kg\per\meter\per\second} \\
	$\lambda$ & thermal conductivity, \qty{}{\watt\per\meter\per\kelvin} \\
	$\rho$ & density, \qty{}{\kg\per\cubic\meter} \\
\end{tabular}

\subsection*{Superscripts and Subscripts}
\begin{tabular}{@{}p{2cm}l}
	av & average \\
	base & base value \\
	char & charging \\
	eff & effective \\
	exp & experiment \\
	fin & fin \\
	l & liquid \\
	loss & loss \\
	$\mathrm{NaNO}_{3}$ & $\mathrm{NaNO}_{3}$ \\
	o & outer \\
	PC & phase change \\
	ref & reference case \\
	s & solid \\
	sim & simulation \\
	tube & tube \\    
\end{tabular}

\vspace{-0.25cm}

\section*{REFERENCES}

\printbibliography[heading=none]

\section*{ACKNOWLEDGEMENT}
Funding by the German Research Foundation within the Priority Program 2403: ‘Carnot-Batteries: Inverse Design from Markets to Molecules’ under project number 525974553 is gratefully acknowledged.

Thanks are also expressed to two Masters students, Arthur Rudek and Dominik Hering, whose work served as the basis for the development of the models presented in this study.

\end{document}